\documentstyle[12pt]{article}
\catcode`\@=11

\@addtoreset{equation}{section}
\def\theequation{\thesection\arabic{equation}}

\def\@normalsize{\@setsize\normalsize{15pt}\xiipt\@xiipt
\abovedisplayskip 14pt plus3pt minus3pt%
\belowdisplayskip \abovedisplayskip
\abovedisplayshortskip  \z@ plus3pt%
\belowdisplayshortskip  7pt plus3.5pt minus0pt}
\def\small{\@setsize\small{13.6pt}\xipt\@xipt
\abovedisplayskip 13pt plus3pt minus3pt%
\belowdisplayskip \abovedisplayskip
\abovedisplayshortskip  \z@ plus3pt%
\belowdisplayshortskip  7pt plus3.5pt minus0pt
\def\@listi{\parsep 4.5pt plus 2pt minus 1pt
            \itemsep \parsep
            \topsep 9pt plus 3pt minus 3pt}}

\def\underline#1{\relax\ifmmode\@@underline#1\else
        $\@@underline{\hbox{#1}}$\relax\fi}
\@twosidetrue
\relax

\catcode`@=12

\catcode`\@=11

\def\section{\@startsection{section}{1}{\z@}{3.5ex plus 1ex minus
   .2ex}{2.3ex plus .2ex}{\large\bf}}
\def\thesection{\arabic{section}.}

\def\ps@headings{\def\@oddfoot{}\def\@evenfoot{}
\def\@oddhead{\hbox{}\hfill
        \makebox[.5\textwidth]{\raggedright\ignorespaces --\thepage{}--
        \hfill }}
\def\@evenhead{\@oddhead}
\def\subsectionmark##1{\markboth{##1}{}}
}

\ps@headings

\catcode`\@=12

\relax

\def\figcap{\section*{Figure Captions\markboth
        {FIGURECAPTIONS}{FIGURECAPTIONS}}\list
        {Fig. \arabic{enumi}:\hfill}{\settowidth\labelwidth{Fig. 999:}
        \leftmargin\labelwidth
        \advance\leftmargin\labelsep\usecounter{enumi}}}
 \relax
\def\tablecap{\section*{Table Captions\markboth
        {TABLECAPTIONS}{TABLECAPTIONS}}\list
        {Table \arabic{enumi}:\hfill}{\settowidth\labelwidth{Table 999:}
        \leftmargin\labelwidth
        \advance\leftmargin\labelsep\usecounter{enumi}}}
 \relax
\def\reflist{\section*{References\markboth
        {REFLIST}{REFLIST}}\list
        {[\arabic{enumi}]\hfill}{\settowidth\labelwidth{[999]}
        \leftmargin\labelwidth
        \advance\leftmargin\labelsep\usecounter{enumi}}}
 \relax

\catcode`\@=11

\def\marginnote#1{}
\newcount\hour
\newcount\minute
\newtoks\amorpm
\hour=\time\divide\hour by60
\minute=\time{\multiply\hour by60 \global\advance\minute by-
\hour}
\edef\standardtime{{\ifnum\hour<12 \global\amorpm={am}%
    \else\global\amorpm={pm}\advance\hour by-12 \fi
    \ifnum\hour=0 \hour=12 \fi
    \number\hour:\ifnum\minute<100\fi\number\minute\the\amorpm}}
\edef\militarytime{\number\hour:\ifnum\minute<100\fi\number\minute}
\def\draftlabel#1{{\@bsphack\if@filesw {\let\thepage\relax
  \xdef\@gtempa{\write\@auxout{\string
    \newlabel{#1}{{\@currentlabel}{\thepage}}}}}\@gtempa
    \if@nobreak \ifvmode\nobreak\fi\fi\fi\@esphack}
     \gdef\@eqnlabel{#1}}
\def\@eqnlabel{}
\def\@vacuum{}
\def\draftmarginnote#1{\marginpar{\raggedright\scriptsize\tt#1}}
\def\draft{\oddsidemargin -.5truein
        \def\@oddfoot{\sl preliminary draft \hfil
        \rm\thepage\hfil\sl\today\quad\militarytime}
        \let\@evenfoot\@oddfoot \overfullrule 3pt
        \let\label=\draftlabel
        \let\marginnote=\draftmarginnote
   
\def\@eqnnum{(\theequation)\rlap{\kern\marginparsep\tt\@eqnlabel}%
\global\let\@eqnlabel\@vacuum}  }
\def\preprint{\twocolumn\sloppy\flushbottom\parindent 1em
        \leftmargini 2em\leftmarginv .5em\leftmarginvi .5em
        \oddsidemargin -.5in    \evensidemargin -.5in
        \columnsep 15mm \footheight 0pt
        \textwidth 250mmin      \topmargin  -.4in
        \headheight 12pt \topskip .4in
        \textheight 175mm
        \footskip 0pt
        
\def\@oddhead{\thepage\hfil\addtocounter{page}{1}\thepage}
        \let\@evenhead\@oddhead \def\@oddfoot{} \def\@evenfoot{} 
}
\def\titlepage{\@restonecolfalse\if@twocolumn\@restonecoltrue\onecolumn
     \else \newpage \fi \thispagestyle{empty}\c@page\z@
        \def\thefootnote{\fnsymbol{footnote}} }
\def\endtitlepage{\if@restonecol\twocolumn \else  \fi
        \def\thefootnote{\arabic{footnote}}
        \setcounter{footnote}{0}}  
\catcode`@=12
\relax

\def\ps@headings{\def\@oddfoot{}\def\@evenfoot{}
\def\@oddhead{\hbox{}\hfill
        \makebox[.5\textwidth]{\raggedright\ignorespaces --\thepage{}--
        \hfill }}
\def\@evenhead{\@oddhead}
\def\subsectionmark##1{\markboth{##1}{}}
}

\ps@headings

\relax

\def\firstpage#1#2#3#4#5#6{
\begin{document}
\begin{titlepage}
\nopagebreak
\title{\begin{flushright}
        \vspace*{-1.8in}
        {\normalsize CERN-TH/97-365}\\[-9mm]
        {\normalsize hep-th/9712164}\\[4mm]
\end{flushright}
\vspace{2cm}
{#3}}
\author{\large #4 \\[0.0cm] #5}
\maketitle
\vskip 2mm
\nopagebreak 
\begin{abstract}
{\noindent #6}
\end{abstract}
\vfill
\begin{flushleft}
\rule{16.1cm}{0.2mm}\\[-3mm]
$^{\star}${\small Research supported in part by\vspace{0mm}
 the EEC under TMR contract ERBFMRX-CT96-0090.
\\[-3mm]
$^\dag$e-mail: Alexandros.Kehagias@cern.ch \\[-3mm]
\hspace{1.45cm}Herve.Partouche@cern.ch } \\[0mm]
CERN-TH-97-365\\[-3mm]
December 1997
\end{flushleft}
\thispagestyle{empty}
\end{titlepage}}

\def\simlt{\stackrel{<}{{}_\sim}}
\def\simgt{\stackrel{>}{{}_\sim}}
\newcommand{\dal}{\raisebox{0.085cm}
{\fbox{\rule{0cm}{0.07cm}\,}}}

\newcommand{\be}{\begin{eqnarray}}
\newcommand{\ee}{\end{eqnarray}}
\newcommand{\btau}{\bar{\tau}}
\newcommand{\p}{\partial}
\newcommand{\bp}{\bar{\partial}}
\newcommand{\cR}{{\cal R}}
\newcommand{\tR}{\tilde{R}}
\newcommand{\tcR}{\tilde{\cal R}}
\newcommand{\hR}{\hat{R}}
\newcommand{\hcR}{\hat{\cal R}}
\newcommand{\oE}{\stackrel{\circ}{E}}
\renewcommand{\p}{\partial}
\renewcommand{\bp}{\bar{\partial}}

\newcommand{\gsi}{\,\raisebox{-0.13cm}{$\stackrel{\textstyle
>}{\textstyle\sim}$}\,}
\newcommand{\lsi}{\,\raisebox{-0.13cm}{$\stackrel{\textstyle
<}{\textstyle\sim}$}\,}
\date{}
\firstpage{3118}{IC/95/34}
{\large {\Large D}--{\Large I}NSTANTON {\Large C}ORRECTIONS AS    
(p,q)--{\Large S}TRING {\Large E}FFECTS \\ AND  
{\Large N}ON--{\Large R}ENORMALIZATION  
{\Large T}HEOREMS$^\star$ \\
\phantom{X}}
{A. Kehagias and H. Partouche$^\dag$} 
{
\normalsize\sl Theory Division, CERN, 1211 Geneva 23, Switzerland
}
{We discuss higher derivative  
interactions in the type IIB superstring in ten 
dimensions. 
From the fundamental string point of view, the non-perturbative corrections 
are due to  D-instantons. We argue that they can alternatively be 
understood  as  arising from $(p,q)$-strings. We derive  
a non-renormalization theorem for eight-derivative bosonic  interactions, 
which states that terms involving either NS-NS or R-R fields occur at
tree-level and one-loop only. 
By using the $SL(2,{\bf Z})$ symmetry of M-theory on $T^2$, 
we show that in order for  the possible $R^{3m+1}$ $(m=1,2,...)$ 
interactions in M-theory to have a  consistent perturbative 
expansion in nine dimensions, $m$ must be odd. Thus, 
only   $R^{6N+4}$ $(N=0,1,...)$ terms can be present in M-theory 
and their string theory counterparts arise at $N$ and $2N+1$ loops. 
Finally, we treat an example of fermionic term. 
}
\section{Introduction}

One of the lessons of the recent developments in string theory is that there
is a multitude of theories, whose fundamental objects are of different 
nature and which however give rise to completely equivalent physics. At the
present time, all ten-dimensional superstring theories are related to one 
another either after being compactified to lower dimensions or being 
interpreted as different limits of more 
fundamental theories such as M- or F- theory 
in eleven and twelve dimensions, respectively. 
In this paper, we will concentrate on the 
type IIB string and we will show how the standard 
interpretation of its effective
action in terms of perturbative string and non-perturbative instanton 
corrections can alternatively be interpreted as arising from 
the $(p,q)$-strings. In general, besides these $(p,q)$-strings \cite{S1},
the various type IIB branes consist in solitons, which are  
the self-dual
three-brane, an $SL(2,{\bf Z})$ multiplet of five-branes,  the 
seven-brane and finally the D-instanton. 
These $p$-branes  play an 
important role in the theory in dimensions lower than 
or equal to $10-(p+1)$ since
only in these cases can their Euclidean $(p+1)$-world volume  wrap around 
cycles of the compact space and contribute  to the effective 
action as standard instantons. However, 
since we will discuss here the type IIB
superstring in dimension nine or ten only, it is natural to think that 
beside string loop calculations, the non-perturbative corrections can only 
arise from D-instantons. This is in fact precisely what was shown in \cite{GG}.
Since the D-instanton  breaks half of the supersymmetries \cite{GGP}, 
the sixteen fermionic zero modes, which are generated by the 
instanton background,   can be saturated by eight derivatives. 
These interactions are of order ${\alpha'}^3$ and some of them
have been evaluated  either in $\sigma$-model perturbation theory \cite{G} or 
by a direct string-amplitude calculation \cite{S2}--\cite{ST}. 
Besides the perturbative 
corrections,  the form of  the non-perturbative ones, which 
arise from multiply-charged one-instantons
(anti-instantons)  backgrounds, for the $R^4$ 
term, has been discussed 
in \cite{GG} and a non-renormalization theorem has been proved in \cite{B}.
This particular term has also 
been conjectured to exist in 
the eleven-dimensional M-theory \cite{GGV,GV}
and compactifications of the latter give 
results consistent with string-theory expectations \cite{RT}--\cite{PKi}. 
The $SL(2,{\bf Z})$ 
invariant four-point amplitude has been worked out in \cite{R}.

The type IIB supergravity has a $U(1)$ symmetry that rotates the two 
supercharges \cite{S}. 
Although the $R^4$ term is neutral under this $U(1)$, there are 
other  eight-derivative interactions that have a non-zero charge.  
The form of the non-perturbative corrections
to such charged terms  has  been proposed in \cite{PK}.  More precisely,
the tree-level four-point 
effective action at order ${\alpha'}^3$, which contains 
all the fields in the NS-NS sector, has been 
completed in an $SL(2,{\bf Z})$-invariant way. The resulting  action involves 
the R-R fields (except the four-form  which is $SL(2,{\bf Z})$-invariant) 
and it is consistent with the D-instanton interpretation.  

Here, we derive the form of the four-point 
effective action of \cite{PK} and extend it to other eight derivative
interactions. This derivation is based on the fact  that the tree-level 
eight-derivative effective action in the NS-NS sector
is invariant under a subgroup $\Gamma_{\infty}$ of $SL(2,{\bf Z})$.
Therefore, $SL(2,{\bf Z})$ invariance 
must be achieved by taking the orbit of each 
term under  $\Gamma_{\infty}\backslash SL(2,{\bf Z})$ 
whose elements are in one-to-one
correspondence with the pairs $p,q$ of integers whose g.c.d. is $1$. 
In this case, the perturbative as well as the non-perturbative corrections 
can alternatively be viewed as arising from $(p,q)$-strings rather than  
fundamental string loops and D-instantons.  
 
In should be noted that although we have started from the tree-level 
eight derivative interactions of the NS-NS fields only, we expect
from the symmetries that there also exist interactions involving $2r$
($r=1,...,4)$ R-R fields and arising at tree level. Their precise
normalization should be evaluated  by a string calculation and their orbit 
under $\Gamma_{\infty}\backslash SL(2,{\bf Z})$ is then completely determined. 
By showing that S-duality relates $n$-loop  to $(1-n)$-loop corrections 
for the eight-derivative interactions,
we obtain a non-renormalization theorem  in type IIB theory.
 According to this, 
an eight-derivative  term 
involving either NS-NS or R-R fields has only tree level and one-loop 
perturbative  corrections.     

The  method  used for the 
determination of the form of the eight-derivative interactions can also be 
applied to determine the perturbative and non-perturbative corrections of 
other terms as well.  
We discuss as an example   the $\lambda^{16}$ 
fermionic interaction recently considered in the literature, where
$\lambda$ is the complex dilatino. Our result is 
in   perfect agreement with the M-theory analysis 
of \cite{GGK}. We also consider $6m+2$-derivative interactions of the 
form $R^{3m+1}$ which are terms consistent with eleven-dimensional Lorentz 
invariance in M-theory \cite{RT}. By applying the $SL(2,{\bf Z})$ 
symmetry in nine dimensions \cite{HT}, we 
find that the  allowed interactions  are of the form $R^{6N+4}$ 
$(N=0,1,...)$ and that they occur at $N$ and $2N+1$ loops only.   

In the next section, we recall some known results for  the type IIB theory.
In section 3 we employ the S-duality symmetry in order to determine 
the perturbative and non-perturbative corrections to  eight-derivative 
interactions.
In section 4 we derive a  non-renormalization theorem for the eight-derivative 
bosonic interactions and in section 5 we consider 
fermionic as well as $R^{3m+1}$ terms. 
Finally, in Appendix A we explicitly 
give the $SL(2,{\bf Z})$ invariant effective action derived from the 
tree-level NS-NS sector only and in Appendix B we present some
properties of the non-holomorphic modular forms we are using.

\section{Tree-level NS-NS sector}

The massless  bosonic spectrum of  the type IIB superstring theory 
consists in  the graviton $g_{MN}$, 
the dilaton $\phi$ and the antisymmetric tensor $B^1_{MN}$ 
in the NS-NS sector, while in the R-R sector it contains  
the axion  $\chi$, the two-form $B^2_{MN}$ and the 
self-dual four-form field $A_{MNPQ}$. 
The fermionic superpartners are a complex Weyl gravitino $\psi_M$
and a  complex Weyl dilatino $\lambda$.
The theory has two 
supersymmetries generated by two supercharges of the same chirality.   
It has in addition a conserved $U(1)$ 
charge, which generates  rotations of the two 
supersymmetries and under  which some of the  fields  are charged 
\cite{S}. 
In particular, 
the graviton and the four-form field are neutral,  the antisymmetric  
tensors have charge  $q=1$, the  scalars have $q=2$, whereas 
the gravitino and the dilatino have charges $q=1/2$ and $q=3/2$, respectively.

The two scalars of 
the theory can be combined into a complex 
one, $\tau=\tau_1+i\tau_2$, defined by
\be
\tau=\chi+ie^{-\phi}\, , \label{tau}
\ee
which parametrizes an $SL(2,{\bf R})/U(1)$ coset space.
At lowest order in 
$\alpha'$, the bosonic effective action of the type IIB string theory 
in the Einstein frame  is 
written as
\be
{\cal{S}}_{3pt}= \frac{1}{2\kappa_{10}^2}\int d^{10}x \sqrt{-g} 
\left[R-\frac{1}{2\tau_2^2}\partial_M\tau\partial^M\btau
-\frac{1}
{12\tau_2}(\tau H^1+H^2)_{KMN}(\btau H^1+H^2)^{KMN}\right] , 
\label{IIB}
\ee
where $H^\alpha_{KMN}=\p_KB^\alpha_{MN}+\mbox{cyclic}$ for $\alpha=1,2$,
$\kappa_{10}^2=2^6\pi^7{\alpha'}^4$ 
and we have set the four-form to zero.
It has an  $SL(2,{\bf R})$ symmetry that acts as 
\be
\tau\rightarrow \frac{a\tau+b}{c\tau+d}\, , ~~~~
B^\alpha_{MN}\rightarrow {(\Lambda^T)^{-1}}^\alpha_{\mbox{\phantom{a}}\beta} 
B^\beta_{MN} 
\, , ~~~g_{MN}\rightarrow g_{MN}\, , ~~~ \Lambda= \left(\matrix{a & b 
\cr c& d}\right) \!\in\! SL(2,{\bf R})\, , \label{ss}
\ee
while  the fermions transform accordingly as 
\be
\psi_M\rightarrow \left(\frac{c\btau+d}{c\tau+d}\right)^{1/4}\psi_M,~~~
\lambda\rightarrow
\left(\frac{c\btau+d}{c\tau+d}\right)^{3/4}\lambda. \label{psi} 
\ee

The action (\ref{IIB}) can be written in the string frame,
\be
{\cal{S}}_{3pt}= \frac{1}{2\kappa_{10}^2}\int d^{10}x \sqrt{-G} 
\left\{e^{-2\phi}\left[{\cal R}\!-\!{1\over 2}(\p
  \phi)^2\!-\!{1\over 12}(H^1)^2 \right]\!\!- \!\!\left[{1\over 2}(\p \chi)^2\!+\!{1\over 12}(\chi
  H^1+H^2)^2 \right]\right\} ,
\label{IIBs}
\ee
where $G_{MN}=e^{\phi/2}g_{MN}$ and ${\cal R}_{MNPQ}$ are the metric and
Riemann tensor in this frame, respectively. From this, we see that the tree level
NS-NS sector kinetic terms have a $e^{-2\phi}$ normalization. In order
for the R-R fields to have the same tree level normalization, we have
to rescale the R-R fields as
\be
\chi\to \chi'=e^\phi \chi\quad , \qquad B^2_{MN}\to
B^{2'}_{MN}=e^\phi B^2_{MN}\, .\label{rescaling}
\ee

The next-to-leading order corrections to the tree-level 
effective action for the NS-NS sector can be  
calculated either in $\sigma$-model perturbation theory or by string 
amplitudes.  At the four-point level,  they are written as
\begin{eqnarray}
{\cal{S}}_{4pt}^{tree}\!\!\!\!&=&\!\!\!\!\frac{1}{2\kappa_{10}^2}
\int\!\!
d^{10}x\sqrt{-g}\left[\frac{{\alpha'}^3}{3\cdot 2^6}\zeta(3)\tau_2^{3/2}
\Big{(}t_8^{ABCDEFGH}t_8^{MNPQRSTU}\!\!+\!
\frac{1}{8}\varepsilon_{10}^{ABCDEFGHIJ}{\varepsilon_{10}^{MNPQRSTU}}_{IJ}
\Big{)}\right. \nonumber  \\
&&\left.
\times \tR_{ABMN}\tR_{CDPQ}\tR_{EFRS}\tR_{GHTU}
\phantom{ \frac{{\alpha}^3}{3^4}}\!\!\!\!\!\!\!\!
\right]\, , \label{R4}
\end{eqnarray}
where  
\be
{\tR_{MN}}^{\mbox{\phantom{PQ}}PQ}={R_{MN}}^{PQ}+
e^{-\phi/2}\nabla_{[M}{H^1_{N]}}^{PQ}-
{g_{[M}}^{[P}\nabla_{N]}\nabla^{Q]}\phi ~~,  \label{hR}
\ee
and we use the convention  $T_{[MN]}=\frac{1}{2}(T_{MN}-T_{NM})$.
The tensor $t_8$ is defined in \cite{S2} in terms of $\eta^{MN}$ and 
$\varepsilon_{10}$ is the totally
antisymmetric symbol in ten dimensions. However, by general
covariance,  $t_8$ should be written in terms
of the metric $g^{MN}$ and the curved space analogue of  
$\varepsilon_{10}\varepsilon_{10}$ should be  understood. 

As has been noted in \cite{GS}, $\tR_{MNPQ}$ is the linearized Riemann tensor
constructed from the connection
\be
\hat{\omega}_{\bar{A}\bar{B}M}={\omega}_{\bar{A}\bar{B}M}
+\frac{1}{2}e^{-\phi/2}
{H^1}_{\bar{A}\bar{B}M}+\frac{1}{4}\left(\delta_{\bar{B}M}\partial_{\bar{A}}
\phi-\delta_{\bar{A}M}\partial_{\bar{B}}
\phi\right)\, ,  
\ee
where $\bar{A},\bar{B}$ are flat space indices. 
The  generalized Riemann tensor 
defined by
\be
{\hat{R}}^{\bar{A}}_{\phantom{A}\bar{B}MN}=
\partial_M{\hat{\omega}^{\bar{A}}}_{\phantom{A}\bar{B}N}-
\partial_N{\hat{\omega}^{\bar{A}}}_{\phantom{A}\bar{B}M}+ 
{\hat{\omega}^{\bar{A}}}_{\phantom{A}\bar{C}M}
{\hat{\omega}^{\bar{C}}}_{\phantom{C}\bar{B}N}
- {\hat{\omega}^{\bar{A}}}_{\phantom{A}\bar{C}N}
{\hat{\omega}^{\bar{C}}}_{\phantom{C}\bar{B}M}  
\ee
is then 
\begin{eqnarray}
{\hat{R}_{MN}}^{\mbox{\phantom{PQ}}PQ}&=&{R_{MN}}^{PQ}+
e^{-\phi/2}\nabla_{[M}{H^1_{N]}}^{PQ}-
{g_{[M}}^{[P}\nabla_{N]}\nabla^{Q]}\phi\nonumber \\&&
-e^{-\phi}{H^1_{[M}}^{C[P}{H^1_{N]C}}^{Q]}+
\frac{1}{4}{g_{[M}}^{[P}\p_{N]}\phi
\p^{Q]}\phi
-\frac{1}{2}{g_{[M}}^{[P}{g_{N]}}^{Q]}\p_K\phi\p^K\phi\label{HR}\\&&
-\frac{1}{2}e^{-\phi/2}\p^{[P}\phi{{H^1}^{Q]}}_{MN}-\frac{1}{2}e^{-\phi/2}
\partial_{[M}\phi {H^1_{N]}}^{PQ} -\frac{1}{2}e^{-\phi/2}{g_{[M}}^{[P}
{H^1_{N]}}^{Q]C}\partial_C\phi\, , \nonumber 
\end{eqnarray}
and the full tree-level 
interactions at order ${\alpha'}^3$ take the form 
\begin{eqnarray}
{\cal{S}}^{tree}\!\!\!\!\!&=&\!\!\!\!\!\frac{1}{2\kappa_{10}^2}
\int\!\! 
d^{10}x\sqrt{-g}\!\left[\!\frac{{\alpha'}^3}{3\cdot 2^6}\zeta(3)\tau_2^{3/2}
\!\Big{(}t_8^{ABCDEFGH}t_8^{MNPQRSTU}\!\!\!+\!
\frac{1}{8}\varepsilon_{10}^{ABCDEFGHIJ}\!{\varepsilon_{10}^{MNPQRSTU}}_{IJ}
\!\Big{)}\right. \nonumber  \\
&&\left.
\times \hat{R}_{ABMN}\hat{R}_{CDPQ}\hat{R}_{EFRS}\hat{R}_{GHTU} 
\phantom{\frac{{\alpha}^3}{ 2^6}}\!\!\!\!\!\!\!
\right]\, . \label{R4f}
\end{eqnarray}

\section{Non-perturbative corrections from S-duality}

We will now proceed  to  determine perturbative and non-perturbative 
corrections to the tree-level NS-NS sector at ${\alpha'}^3$ order of the type 
IIB superstring theory.  This will be worked out  by employing the 
S-duality symmetry of the theory, which relates the NS-NS fields to the R-R 
ones.

It is easily seen from (\ref{ss}) that  $\nabla_M H^1_{NPQ}$ and
 $\nabla_M\partial_N\phi$ do not transform covariantly under 
$SL(2,{\bf R})$. 
It is convenient to  introduce
$SL(2,{\bf R})$-covariant objects and, for this, let us 
recall that 
the complex scalar $\tau$ parametrizes a $SL(2,{\bf R})/U(1)$ coset space. 
In general, the group 
$SL(2,\bf{R})$ can be represented by a matrix $V^\alpha_{\pm}$ 
\cite{S,GG}:
\be
 V=\left(\matrix{V^1_- & V^1_+\cr 
V^2_- & V^2_+ }\right)= \frac{1}{\sqrt{-2 i\tau_2}} 
\left(\matrix{\btau e^{-i\theta} & \tau e^{i\theta} \cr 
e^{-i\theta} & e^{i\theta} }\right)\, . 
\ee
 The local $U(1)$ is realized by the shift 
$\theta\rightarrow \theta+\Delta\theta$ and the global 
$SL(2,\bf{R})$ acts from the left. One may define the quantities 
\be
P_M=-\epsilon_{\alpha\beta}V^\alpha_+\p_MV^\beta_+=ie^{2i\theta}
\frac{\p_M\tau}{2\tau_2}\, , ~~~ Q_M=-i\epsilon_{\alpha\beta}V^\alpha_+
\p_MV^\beta_-=\p_M\theta -
\frac{\p_M\tau_1}{2\tau_2}\, ,   \label{PQ} 
\ee
where $Q_M$ is a composite $U(1)$ gauge connection and $P_M$ has charge 
$q=2$. We also define the complex three-form
\be
G_{KMN}=-\sqrt{2i}\delta_{\alpha\beta}
V^\alpha_+H^\beta_{KMN}=-i\frac{e^{i\theta}}{\sqrt{\tau_2}}
(\tau H^1_{KMN}+H^2_{KMN})\, , \label{G}
\ee
with charge $q=1$. We  fix the $U(1)$ gauge by choosing  
$\theta\equiv 0$ from now on. 
In this case, the global $SL(2,\bf{R})$ transformation is 
non-linearly realized and the various quantities in eqs.(\ref{PQ}) and 
(\ref{G}) transform as
\be
P_M\rightarrow \frac{c\btau+d}{c\tau+d}P_M\, , ~~ Q_M\rightarrow 
Q_M+\frac{1}{2i}\p_M\ln \left(\frac{c\btau+d}{c\tau+d}\right)\,  ,~~
G_{KMN}\rightarrow \left(\frac{c\btau+d}{c\tau+d}\right)^{1/2}\!\!G_{KMN}
\, . \label{tra}
\ee
We also define  the covariant derivative $D_M=\nabla_M-iqQ_M$, 
which transforms under $SL(2,\bf{R})$ as  
\be
D_M\rightarrow \left(\frac{c\btau+d}{c\tau+d}\right)^{q/2}D_M \, . \label{D} 
\ee

The generalized Riemann tensor  $\hat{R}_{MNPQ}$ in eq.(\ref{HR}) 
can now be written in terms of the previous 
covariant quantities as
\begin{eqnarray}
{{\hat{R}_{MN}}^{\mbox{\phantom{PQ}}PQ}}&=&
\frac{1}{2}{R_{MN}}^{PQ}+\frac{1}{2}\left( D_{[M}{G_{N]}}^{PQ}+
P_{[M}\bar{G}_{N]}^{\mbox{\phantom{N}}PQ} \right)
-{g_{[M}}^{[P}D_{N]}P^{Q]}-\frac{3}{4}{g_{[M}}^{[P}
P_{N]}P^{Q]}\nonumber \\
&&-\frac{1}{2}{g_{[M}}^{[P}{g_{N]}}^{Q]}P_KP^K
+\frac{5}{4}{g_{[M}}^{[P}P_{N]}\bar{P}^{Q]}-\frac{1}{4}
{G_{[M}}^{C[P}{G_{N]C}}^{Q]}
-\frac{1}{4}
{G_{[M}}^{C[P}{\bar{G}_{N]C}}^{\mbox{\phantom{PQ}}Q]}\nonumber \\
&&
-\frac{1}{2}{g_{[M}}^{[P}{g_{N]}}^{Q]}P_K{\bar{P}}^K
-\frac{1}{4}{g_{[M}}^{[P}{G_{N]}}^{Q]C}P_C
-\frac{1}{4}{g_{[M}}^{[P}{\bar{G}_{N]}}^{\phantom{N]}Q]C}
P_C \label{hhhR} \\ 
&&
+\frac{1}{4}\left({G_{MN}}^{[P}P^{Q]}+{G^{PQ}}_{[M}P_{N]}\right)
+\frac{1}{4}\left({{\bar{G}}_{MN}}^{\mbox{\phantom{PQ}}[P}P^{Q]}+
{{\bar{G}}^{PQ}}_{\mbox{\phantom{....}}[M}P_{N]}\right) 
+\mbox{c.c.}\;. \nonumber 
\end{eqnarray}
In the first line of the 
r.h.s. of (\ref{hhhR}), we have kept two terms in parenthesis
 since this combination involves the $SL(2,{\bf R})$-covariant quantity 
\be
{\cal D}_{M}G_{NPQ}\equiv D_M G_{NPQ}+P_M\bar{G}_{NPQ}={-i\over \sqrt{\tau_2}}
(\tau \nabla_M H^1_{NPQ}+\nabla_M H^2_{NPQ})\, , 
\ee
which satisfies the Bianchi identity 
\be
{\cal D}_{[M}G_{N]PQ}=-{\cal D}_{[P}G_{Q]MN}\, .
\ee

By using the expression (\ref{hhhR}) for $\hat{R}_{MNPQ}$ in the 
action (\ref{R4f}),
it is clear that the latter is only a part of the full S-duality-invariant 
effective action. $SL(2,{\bf Z})$ transformations generate new terms 
and the S-duality invariant result must contain the orbit of
the $\hat{R}^4$ terms 
under $SL(2,{\bf Z})$. The action (\ref{R4f}) is invariant
under the subgroup 
\be
\Gamma_\infty=\left\{ \left(\matrix{\pm 1&  *\cr 0&\pm 1}\right)\in 
SL(2,{\bf Z})\right\}\, ,
\ee
and thus the $SL(2,{\bf Z})$ orbit  is obtained by summing over all 
matrices $\gamma\in \Gamma_\infty\backslash SL(2,{\bf Z})$. 
By noticing that 
\be
\left(\matrix{\pm 1&  *\cr 0&\pm 1}\right)
\left(\matrix{a&b\cr c&d}\right)=\pm\left(\matrix{*&  *\cr c&d}\right)\, ,
\ee
and that a matrix belongs to $SL(2,{\bf Z})$ if and only if the g.c.d. of its 
second row is $1$, we find that 
a representative 
of the coset $\Gamma_\infty\backslash SL(2,{\bf Z})$ is 
\be
\gamma=\pm\left(\matrix{*&  *\cr p & q}\right)\, , \label{gamma}
\ee
where the g.c.d. $(p,q)=1$ \cite{TT}. The effective action (\ref{R4f}) 
must then be completed to  
\begin{eqnarray}
{\cal{S}}&=&\frac{1}{2\kappa_{10}^2}
\int d^{10}x\sqrt{-g}\left\{\frac{{\alpha'}^3}{3\cdot 2^6}
\Big{(}t_8t_8+
\frac{1}{8}\varepsilon_{10}\varepsilon_{10}
\Big{)}\zeta(3)\frac{1}{2} \sum_{(p,q)=1}\frac{\tau_2^{3/2}}{|p\tau+q|^3}
\right. \nonumber \\ 
&&
\left[\frac{1}{2}{R_{MN}}^{PQ}+\frac{1}{2}
\left(\frac{p\btau+q}{p\tau+q}\right)^{1/2}\left( D_{[M}{G_{N]}}^{PQ}+
P_{[M}\bar{G}_{N]}^{\phantom{N]}PQ}\right)
-\left(\frac{p\btau+q}{p\tau+q}\right){g_{[M}}^{[P}D_{N]}P^{Q]}\right.
\nonumber \\&&
-\frac{3}{4}\left(\frac{p\btau+q}{p\tau+q}\right)^2{g_{[M}}^{[P}
P_{N]}P^{Q]}-\frac{1}{2}\left(\frac{p\btau+q}{p\tau+q}\right)^2
{g_{[M}}^{[P}{g_{N]}}^{Q]}P_KP^K
+\frac{5}{4}{g_{[M}}^{[P}P_{N]}\bar{P}^{Q]}\nonumber \\ &&
-\frac{1}{4}
\left(\frac{p\btau+q}{p\tau+q}\right)
{G_{[M}}^{C[P}{G_{N]C}}^{Q]}
-\frac{1}{4}
{G_{[M}}^{C[P}{\bar{G}_{N]C}}^{\mbox{\phantom{PQ}}Q]}
-\frac{1}{2}{g_{[M}}^{[P}{g_{N]}}^{Q]}P_K{\bar{P}}^K\label{sss} \\ 
&&
-\frac{1}{4}\left(\frac{p\btau+q}{p\tau+q}\right)^{1/2}
{g_{[M}}^{[P}{\bar{G}_{N]}}^{\phantom{N]}Q]C}P_C 
+\frac{1}{4}\left(\frac{p\btau+q}{p\tau+q}\right)^{3/2}
\left({G_{MN}}^{[P}P^{Q]}+{G^{PQ}}_{[M}P_{N]}\right)\nonumber \\
&&\left.\left.-\frac{1}{4}\left(\frac{p\btau+q}{p\tau+q}\right)^{3/2}
g_{[M}^{\phantom{[M}[P}{G_{N]}}^{Q]C}P_C
+\frac{1}{4}\left(\frac{p\btau+q}{p\tau+q}\right)^{1/2}
\left({{\bar{G}}_{MN}}^{\mbox{\phantom{PQ}}[P}P^{Q]}+
{{\bar{G}}^{PQ}}_{\mbox{\phantom{....}}[M}P_{N]}\right) +\mbox{c.c.}
\right]^4\right\}, 
\nonumber 
\end{eqnarray}
where the factor $1/2$ in front of the discrete sum 
is due to the sign ambiguity of eq.(\ref{gamma}). Then the sum over 
$p,q$ with $(p,q)=1$ can be extended in an ordinary sum by replacing 
\be
\zeta(3)\sum_{(p,q)=1}\to {\sum_{p,q}}' \, ,~~~~~~~~
 \mbox{with } p,q \in {\bf Z}^2-\{0,0\}\, .
\ee
We may expand the fourth power in the square brackets in eq.(\ref{sss}) in 
terms of $\left(\frac{p\btau+q}{p\tau+q}\right)^{k/2}$
with $k=-16,\dots,16$ and the result is given in Appendix A. Here, 
for simplicity,  let us  consider as an example 
 the contributions that involve  three terms 
in the second line of eq.(\ref{sss}). These terms are the only ones 
that contribute to the four-point string amplitudes.
We find 
\begin{eqnarray}
{\cal{S}}&=&\frac{1}{2\kappa_{10}^2}
\int d^{10}x\sqrt{-g}\left\{ \frac{{\alpha'}^3}{3\cdot 2^7}
\Big{(}t_8t_8+
\frac{1}{8}\varepsilon_{10}\varepsilon_{10}
\Big{)}{\sum_{p,q}}'\frac{\tau_2^{3/2}}{|p\tau+q|^3}
\right. \nonumber \\ 
&&
\left.\left[\frac{1}{2}{R_{MN}}^{PQ}+\frac{1}{2}
\left(\frac{p\btau+q}{p\tau+q}\right)^{1/2} D_{[M}{G_{N]}}^{PQ}
-\left(\frac{p\btau+q}{p\tau+q}\right){g_{[M}}^{[P}D_{N]}P^{Q]}+\mbox{c.c.}
\right]^4\right\}
\nonumber \\
&=&\frac{1}{2\kappa_{10}^2}\int d^{10}x\sqrt{-g} 
\left\{\mbox{\phantom{\huge{X}}}\right.\!\!\!\!\!\!\!\!\!\!
\frac{{\alpha'}^3}{3\cdot 2^7}(t_8t_8+\frac{1}{8}\varepsilon_{10}\varepsilon_{10})\times
\nonumber \\ 
&&\left[\mbox{\phantom{\huge{X}}}\right.\!\!\!\!\!\!\!\!\!\!\frac{1}{2}
f_0(\tau,\btau)
\left(\mbox{\phantom{\huge{X}}}\right.\!\!\!\!\!\!\!\!\!\!R^4+
12R^2DP\,D\bar{P}-6RDP\,D\bar{G}^2 
+3R^2 DG\,D\bar{G}\nonumber \\
&&
+6DP^2\,D\bar{P}^2+\frac{3}{8}DG^2\,D\bar{G}^2
+6DP\,D\bar{P}\,DG\,D\bar{G}\!\!\!\!\!\!\!\!
\left.\mbox{\phantom{\huge{X}}}\right)
\nonumber \\
&&+f_1(\tau,\btau)\left(\mbox{\phantom{\huge{X}}}\right.\!\!\!\!\!\!\!\!\!\!
-4R^3DP +\frac{3}{2}R^2DG^2-12RDP^2\,D\bar{P}-6RDP\,DG\,D\bar{G} \label{f} \\
&&
+3DP\,D\bar{P}\,DG^2+\frac{3}{2}DP^2\,D\bar{G}^2
+\frac{1}{4}DG^3\,D\bar{G}\!\!\!\!\!\!\!\!
\left.\mbox{\phantom{\huge{X}}}\right)\nonumber \\
&&+
f_2(\tau,\btau)\left(\mbox{\phantom{\huge{X}}}\right.\!\!\!\!\!\!\!\!\!\!
6R^2DP^2-3RDPDG^2+4DP^3\,D\bar{P}+3DP^2\,DG\,D\bar{G}+\frac{1}{16}DG^4
\left.\mbox{\phantom{\huge{X}}}\!\!\!\!\!\!\!\!\!\!\right) \nonumber \\
&&
+f_3(\tau,\btau)\left(\mbox{\phantom{\huge{X}}}\right.\!\!\!\!\!\!\!\!\!\!
-4RDP^3+\frac{3}{2}DP^2DG^2 \!\!\!\!\!\!\!\!\!\!
\left.\mbox{\phantom{\huge{X}}}\right)
+f_4(\tau,\btau)
DP^4
+\mbox{c.c.} \!\!\!\!\!\!\!\!\!\!
\left.\mbox{\phantom{\huge{X}}}\right] \!\!\!\!\!\!\!\!\!\!
\left.\mbox{\phantom{\huge{X}}}\right\}\, , \nonumber  
\end{eqnarray}
where $DG$ stands for $(DG)_{MNPQ}\equiv D_{M}G_{NPQ}$, $DP$ for 
$(DP)_{MNPQ}\equiv g_{MP}D_NP_Q$ and similarly for $D\bar{G}$ and $D\bar{P}$
of $U(1)$ charge $q=-1,-2$, respectively. This result is in perfect agreement 
with \cite{PK}, where  the S-duality invariant corrections to 
the eight-derivative four-point tree-level interactions of the NS-NS
sector were conjectured.

The functions $f_k(\tau,\btau)$ are defined as 
\be
f_k(\tau,\btau)=
{\sum_{p,q}}'\frac{\tau_2^{3/2}}{(p\tau+q)^{3/2+k}(p\btau+q)^{3/2-k}}\, ,
\label{fk}
\ee
and they transform under  $SL(2,\bf{Z})$ as 
\be
f_k(\tau,\btau)\rightarrow \left(\frac{c\tau+d}{c\btau+d}\right)^k
f_k(\tau,\btau)\, , ~~~ \left(\matrix{a & b\cr c& d}\right)\in SL(2,\bf{Z})\, .
\ee 
By considering the full expansion in eq.(\ref{sss}), the  functions $f_k$ 
which 
appear are $f_{0}(\tau,\btau)$, $f_{1}(\tau,\btau)$,...,  
$f_{8}(\tau,\btau)$ and 
their complex conjugates 
$f_{0}(\tau,\btau)$, $f_{-1}(\tau,\btau)$, ..., $f_{-8}(\tau,\btau)$ only.  
This is due to the fact that all 
$f_{k}(\tau,\btau)$ with  $k$ half-integer vanish, as can be seen from their 
definition (\ref{fk}). There are thus no 
interactions involving an odd number of  two-forms, which is not
a surprise since the transformation $-I_{2\times2} \in SL(2,{\bf Z})$ acts as
\be
-I_{2\times 2}~:\quad \tau\to \tau\; , 
\qquad  B_{MN}^\alpha\to - B_{MN}^\alpha\, \label{-I},  
\ee
and does not allow such terms. 

Perturbatively, the type IIB superstring is also invariant under the
world-sheet parity operator $\Omega$ whose effect on the massless bosonic 
fields is
\be
\Omega~:\quad \tau_1\to -\tau_1\;, \qquad B^1_{MN}\to -B^1_{MN}\;, 
\qquad A_{MNPQ}\to -A_{MNPQ}\, ,
\ee
while the graviton, dilaton and R-R two-form are invariant. In the 
$SL(2,{\bf Z})$-invariant action, this symmetry remains unbroken since, once
translated in terms of covariant quantities, it takes the form
\be
\Omega~:\quad f_k(\tau,\btau)\to f_{-k}(\tau,\btau)\, , \quad P_M\to \bar{P}_M\, ,
\quad G_{MNP}\to -\bar{G}_{MNP}\, , \quad Q_M\to -Q_M\, ,
\ee
which amounts to a complex conjugation of the effective action.

We have determined above perturbative and non-perturbative corrections to
eight-deri\-vative interactions
by considering  
the leading order in their small $\tau_2^{-1}$ 
expansion, which is $2\zeta(3)\tau_2^{3/2}$.
In particular, under $SL(2,{\bf Z})$, the latter generates  perturbative 
corrections at one loop
proportional to $\tau_2^{-1/2}$ only, as can be seen from eq.(\ref{exp}). 
We would like to stress that these 
two perturbative contributions 
are actually on an equal footing. By this we mean that if we had started 
from the $\tau_2^{-1/2}$ contributions, we would have generated the 
$\tau_2^{3/2}$
corrections in a similar way. 
As an example, let us explicitly describe how this works on 
the simplest case, namely the $\tau_2^{-1/2}R^4$ term, which gives
\be
\frac{2\pi^2}{3}\tau_2^{-1/2}R^4\to 
\frac{2\pi^2}{3}{1\over 2}\sum_{(p,q)=1}
\frac{\tau_2^{-1/2}}{|p\tau+q|^{-1}}R^4={\pi^2\over 3}
\frac{1}{\zeta(-1)}{\sum_{p,q}}'
\frac{\tau_2^{-1/2}}{|p\tau+q|^{-1}}R^4\, , \label{ff}
\ee
after considering its orbit under $SL(2,{\bf Z})$. Using the alternative 
expression for $f_k$  given in eq.(\ref{alt}), we get
\be
\frac{2\pi^2}{3}\tau_2^{-1/2}R^4\to
f_0(\tau,\btau)R^4\, , 
\ee
which is exactly what we obtained before by starting from the
$2\zeta(3)\tau_2^{3/2}R^4$ term.

We have seen that $SL(2,{\bf Z})$ relates tree and one-loop contributions.
We can show now that there are no other perturbative corrections to the 
eight-derivative interactions compatible with S-duality. 
We will demonstrate this for the  $R^4$
term and the generalization for interactions of non-zero $U(1)$ charge is 
then straightforward. Let us suppose that there exists the 
$n$-loop correction $\tau_2^{3/2-2n}R^4$ which under the action of 
$SL(2,{\bf Z})$ gives 
\be
\tau_2^{3/2-2n}R^4&\rightarrow& {1\over 2}\sum_{(p,q)=1}
   \frac{\tau_2^{3/2-2n}}{|p\tau+q|^{3-4n}}R^4={1\over 2\zeta(3-4n)}
{\sum_{p,q}}'\frac{\tau_2^{3/2-2n}}{|p\tau+q|^{3-4n}}R^4\nonumber \\&&
={1\over 2\zeta(3-4n)}
f_{3/2-2n,0}(\tau,
\btau)R^4
\ee
where,  in general, 
\be
f_{s,k}{(\tau,\btau)}=
{\sum_{m,n}}'\frac{\tau_2^{s}}{(m\tau+n)^{s+k}(m\btau+n)^{s-k}}\, , \label{fkk}
\ee
are non-holomorphic  modular forms of weights $(s+k,s-k)$. Note that 
$f_{3/2,k}$ is identical to $f_{k}$ introduced before. The  
$f_{s,k}$'s for generic $k$ are relevant for interactions with non-zero $U(1)$ 
charge. 
For $n\geq 1$ we have $s=3/2-2n<0$  and  
the infinite sum does not converge but can be defined by analytic 
continuation \cite{M} as
\be
f_{s,k}(\tau,\btau)=\pi^{2s-1}\frac{\Gamma(1-s+k)}{\Gamma(s+k)}
f_{1-s,k}(\tau,\btau) \, . \label{45}  
\ee
The small $\tau_2^{-1}$ expansion of $f_{3/2-2n,0}$ is then 
\be
f_{3/2-2n,0}(\tau,\btau)&=&\pi^{2-4n} 
{\Gamma(2n-1/2)\over \Gamma(3/2-2n) }\, 
\label{exp2}\\&&
\times 
\left(2\zeta(4n-1)\tau_2^{2n-1/2}+\frac{2\sqrt{\pi}\Gamma(2n-1)\zeta(4n-2)}
{\Gamma(2n-1/2)}\tau_2^{3/2-2n}+\cdots \right)\, , \nonumber  
\ee
where the dots stand for instanton corrections. 
We see from the above expansion that S-duality relates the $n$-loop 
contribution to the $(1-n)$-loop  $\tau_2^{2n-1/2}R^4$ term. We conclude then 
that $n=0$ or $n=1$ are the only perturbative contributions consistent with 
S-duality.

The formal sums we considered in this section
have a physical interpretation. We recall that any contribution 
in the effective action that has been determined so far  
by an explicit string or sigma-model calculation is 
invariant under the abelian group
$\Gamma_\infty$. The invariance under the latter implies that in order 
to achieve the S-duality symmetry, all 
contributions to be added are associated to coprime pairs of integers $p,q$.
This is equivalent to considering them as arising from the $(p,q)$-strings
of the type IIB theory \cite{S1}.  
In fact, it is surprising that $(p,q)$-strings play a role already in 
ten dimensions. One should expect that these objects are  important 
for non-perturbative physics in dimensions lower than or equal to 8,
since their Euclidean world volume can then wrap
around two-dimensional cycles of the compact manifold. 
Here, we see that 
the perturbative as well as the non-perturbative
D-instanton corrections  of
the fundamental $(0,1)$-string have an alternative interpretation 
in ten dimensions. They can 
be viewed as  a sum of only ``perturbative'' contributions from all 
$(p,q)$-strings. Indeed, due to the fact that these objects are the images 
of the fundamental string under S-duality, they appear on equal footing in 
the final result.

\section{A Non-renormalization theorem}

Here, we discuss in more detail the structure of the terms we collected in 
Appendix A. At order ${\alpha'}^3$, the contributions that appear 
involve in the Einstein frame: 
\be
NS-NS&:& \partial g\, , \partial \phi\, , e^{-\phi/2}H^1 \nonumber \\
R-R&:& e^{\phi}\partial\tau_1\, , e^{\phi/2}\left(\tau_1H^1+ H^2\right)\, ,
\ee    
and their first derivatives. To determine their string-loop 
expansion  order, we  consider them in the 
string frame. There we find that the  interactions involve 
the independent terms  
\be 
NS-NS&:& \partial G\, , \partial \phi\, ,H^1 \nonumber \\
R-R&:& \partial\tau_1\, , \left(\tau_1H^1+ H^2\right)\, ,
\ee
where $G_{MN}=e^{\phi/2}g_{MN}$.
By using eq.(\ref{exp}), the general structure of the perturbative 
interactions are of the following form
\be
\left(2\zeta(3)e^{-2\phi}k_1+\frac{2\pi^2}{3}k_2\right)e^{2L\phi}&&
\!\!\!\!\!\!\!\!\!
(\p G)^{n_1}\,(\p \phi)^{n_2}\, (H^1)^{n_3}\,(\p \tau_1)^{r_1}\,
(\tau_1 H^1+H^2)^{r_2}\label{L} \\
&&\!\!\!\!\!\!\!\!\!\times(\p\p G)^{m_1}\,(\p\p \phi)^{m_2}\, 
(\p H^1)^{m_3}\,(\p\p \tau_1)^{s_1}\,
(\tau_1 \p H^1+\p H^2)^{s_2}\, , \nonumber
\ee
where $L=\frac{1}{2}(r_1+r_2+s_1+s_2)$ and 
$k_1,k_2$ are rational coefficients depending on the positive 
integers $n_i,m_i,r_j,s_j$ $(i=1,2,3; j=1,2)$, which satisfy  
\be
\begin{array}{cl}
n_1+n_2+n_3+r_1+r_2+2(m_1+m_2+m_3+s_1+s_2) = 8 &
(= \mbox{number of derivatives})\\
n_3+m_3+r_1+s_1 \;\;\mbox{even}& (\Omega\mbox{ invariance})\\
n_3+m_3+r_2+s_2 \;\;\mbox{even}& (-I_{2\times 2}\mbox{ invariance})\, .
\label{con}
\end{array}
\ee
Notice that these conditions imply that $L=0,1,...,4$. To determine
the loop order this corresponds to, we
have to take into account the rescaling (\ref{rescaling}) of the R-R
fields. As a result, the perturbative interactions (\ref{L}) arise at tree
level and one-loop only. However, 
it should be stressed that $k_1=0$ for the terms  involving a R-R field. 
The reason for this is that when we considered the orbit of the tree-level 
NS-NS sector proportional to $2\zeta(3)\tau_2^{3/2}$ in the Einstein frame,
we generated only terms proportional to $\frac{2\pi^2}{3}\tau_2^{-1/2}$ or 
non-perturbative. However, there is no symmetry that excludes  terms  
involving R-R fields at tree level. Therefore, their presence has to be checked by 
an explicit string calculation and then, their $SL(2,{\bf Z})$ orbit 
has to be added. 
However, after performing this procedure, eqs.(\ref{L},\ref{con}) 
remain valid and the only effect of these terms is to change the values 
of $k_1$ and $k_2$ while the structure of eq.(\ref{L}) remains unaffected.
To show  this explicitly,  let us consider as an example 
 in the Einstein frame  the term 
\be
\Big{(}t_8^{ABCDEFGH}t_8^{MNPQRSTU}&\!\!\!+\!\!\!&
\frac{1}{8}\varepsilon_{10}^{ABCDEFGHIJ}{\varepsilon_{10}^{MNPQRSTU}}_{IJ} 
\Big{)} \label{ex}\\ &&
a\,  \tau_2^{3/2}R_{ABMN}R_{CDPQ}g_{ER}\frac{\p_F\tau_1}{\tau_2}
\frac{\p_S\tau_1}{\tau_2}
g_{GT}\frac{\p_H\tau_1}{\tau_2}\frac{\p_U\tau_1}{\tau_2}\, , \nonumber
\ee
where $a$ is a number to be determined by a string calculation. By a
straightforward calculation  following the steps which led us from 
eq.(\ref{R4f}) to  eq.(\ref{inter}), the orbit under $SL(2,{\bf Z})$ of 
(\ref{ex}) is
\be
&&
\Big{(}t_8^{ABCDEFGH}t_8^{MNPQRSTU}+
\frac{1}{8}\varepsilon_{10}^{ABCDEFGHIJ}{\varepsilon_{10}^{MNPQRSTU}}_{IJ}
\Big{)}\label{tt1} {a \over 2\zeta(3)}R_{A
BMN}R_{CDPQ}\times \nonumber \\  &&\left[
\phantom{\frac{1^1}{1^1}}\right.\!\!\!\!\!f_0(\tau,\btau)
\Big{(}g_{ER}P_FP_Sg_{GT}\bar{P}_H\bar{P}_U
+g_{ER}\bar{P}_FP_Sg_{GT}\bar{P}_HP_U
+g_{ER}\bar{P}_FP_Sg_{GT}P_H\bar{P}_U\Big{)} \nonumber \\
&&-f_2(\tau,\btau)\Big{(}g_{ER}\bar{P}_FP_Sg_{GT}P_HP_U 
+g_{ER}P_F\bar{P}_Sg_{GT}P_H{P}_U 
+g_{ER}{P}_F{P}_Sg_{GT}\bar{P}_H{P}_U   \\&&+
g_{ER}{P}_F{P}_Sg_{GT}P_H\bar{P}_U\Big{)}
+f_4(\tau,\btau)\Big{(}g_{ER}P_F{P}_Sg_{GT}P_HP_U\Big{)}+\mbox{c.c.}\left.
\phantom{\frac{1}{1}}\!\!\!\!\!\right]\nonumber \, .  
\ee
 It can now be seen that the 
interactions induced by the  term (\ref{ex}) are of the form (\ref{L}) 
and actually give corrections to $k_2$ only. 

The previous remarks we demonstrated on the specific 
example of eq.(\ref{ex}) remain valid 
for any other term involving $2r$ R-R fields at tree level
and allowed by the symmetries. 
As a result, we can state a non-renormalization theorem:
{\it For the eight-derivative bosonic interactions of the type IIB theory
which involve either NS-NS or R-R fields (except the self-dual four-form), 
there exist only tree level and one-loop 
perturbative corrections.}

We would like to point out that although  our procedure infer the structure 
of the perturbative sector of the type IIB superstring, the non-perturbative 
one may differ from what we obtained. This is due to the fact 
that one may add ``cusp forms'' to the $f_k$'s of the same weights without 
altering their perturbative expansion. By ``cusp forms'' we mean here 
modular functions of the form 
${\sum'}_{m,n\geq 1} a_{m,n}(\tau_2)e^{2i\pi m\tau}e^{-2i\pi n\btau}$. However,
this possibility does not affect the above non-renormalization theorem.
 
\section{Fermionic and $R^{3m+1}$ terms}

To illustrate the generality of the procedure that we presented for 
the eight-derivative bosonic interactions,
we now discuss fermionic as well as $R^{3m+1}$ terms. 
The fermionic term we will consider is related to  
$t_8t_8R^4$ by supersymmetry
and  is the $\lambda^{16}$ interaction which should arise at tree level. 
This
term is in fact the product of the 16 components of the complex Weyl spinor
whose real and imaginary parts are the two dilatinos of same chirality.
The  perturbative and non-perturbative corrections associated to it are
then determined by taking into account as before its orbit under 
$SL(2,{\bf Z})$ and the transformation property (\ref{psi}),
\be
2b\zeta(3) \tau_2^{3/2}\lambda^{16}&\to& 2b\zeta(3) \frac{1}{2} \sum_{(p,q)=1} 
{\tau_2^{3/2} \over |p\tau +q|^3} \left( {p\btau +q\over p\tau+q} \right)^{12}
\lambda^{16}\nonumber \\
&&=b \, {\sum_{p,q}}'{\tau_2^{3/2} \over 
(p\tau +q)^{3/2+12}(p\btau +q)^{3/2-12}}
\lambda^{16}\\ 
&&=b\, f_{12}(\tau,\btau)\, \lambda^{16} \, ,\nonumber 
\ee
where $b$ is a numerical factor. This result is 
in complete agreement with the M-theory calculation of ref. \cite{GGK}.

Interactions of the form $R^{3m+1}$ of 
${\alpha'}^{3m}$ order with $m=1,2,...$ have 
been conjectured to exist in M-theory
\cite{RT}. They are the only terms that can be decompactified in 
eleven dimensions in a Lorentz-invariant way. In the string frame they 
can arise at $m$ loops 
\be
 \int d^{10}x \sqrt{-G} e^{2\phi(m-1)}\cR^{3m+1} \, , \label{mm}
\ee
where $\cR_{MNPQ}$ is the  
Riemann tensor in this frame. 
Eventually, there are also lower order 
perturbative contributions. 

M-theory on $T^2$ has an $SL(2,{\bf Z})$ symmetry, 
which is the T-duality group of the torus. 
It acts on the  complex structure $\Omega$, which 
can be identified with the type IIB complex scalar $\tau$ in nine dimensions.  
By decompactifying to ten dimensions on the type IIB side and following the 
procedure described in section 3, we find that eq.(\ref{mm}) is promoted to the
$SL(2,{\bf Z})$-invariant term  
\be
\int d^{10}x \sqrt{-g}f_{-m/2,0}(\tau,\btau)
R^{3m+1} \, ,  \label{mmm}
\ee 
where,  $f_{-m/2,0}(\tau,\btau)$ has been defined in eq.(\ref{fkk}).
The small $\tau_2^{-1}$ expansion of $f_{-m/2,0}$ as follows from 
eq.(\ref{exp2}) is then 
\be
f_{-m/2,0}(\tau,\btau)&=&\pi^{-(m+1)}\frac{\Gamma(1+m/2)}{\Gamma(-m/2)}
\label{exp3}\\&&
\times 
\left(2\zeta(2+m)\tau_2^{1+m/2}+\frac{2\sqrt{\pi}\Gamma(1/2+m/2)\zeta(1+m)}
{\Gamma(1+m/2)}\tau_2^{-m/2}+\cdots \right)\, , \nonumber  
\ee
where the dots stand for instanton corrections. From the expansion 
(\ref{exp3}),  
we see that there exist two perturbative contributions at $m$ and 
$(m-1)/2$ loops to the $R^{3m+1}$ terms. Then,
a necessary condition for such terms to exist perturbatively
 is that $m$ be odd. As a result, 
we conclude that: {\em Interactions consistent with eleven-dimensional 
M-theory and $SL(2,{\bf Z})$ symmetry in nine dimensions  
 are of  the form $R^{6N+4}$ $(N=0,1,...)$ and arise at $N$ and $2N+1$ loops.}

\noindent{\bf Acknowledgements} 
 
We are grateful to I. Antoniadis, S. Ferrara, 
M.B. Green, B. Pioline, J.G. Russo, A.A. Tseytlin and P. Vanhove for 
useful discussions.  We also thank N.J. Berkovits for his relevant
remarks about loop counting for R-R fields.

\newpage
\begin{flushleft}
{\large\bf Appendix A}\end{flushleft}
\renewcommand{\theequation}{A.\arabic{equation}}
\renewcommand{\thesection}{A.}
\setcounter{equation}{0}
 
We present here the S-duality invariant interactions of the type IIB
effective action, which have been obtained by considering at tree
level the NS-NS sector only. However,  to obtain 
the full action at ${\alpha'}^3$ order, besides these $\hat{R}^4$
terms, it is necessary to extend 
the analysis  to all  other eight-derivative contributions 
that involve $2r$ R-R fields $(r=1,2,3,4)$ and occur at tree level only.

For convenience, we introduce the following 
quantities $X_i$ $(i=0,...,4)$ whose $U(1)$ charges are well defined, namely
$q=i$,
\begin{eqnarray}
(X_0)_{MNPQ} &\equiv& R_{MNPQ}+{5\over 4}\, g_{MP} 
\left(P_N\bar{P}_Q+\bar{P}_NP_Q\right)-g_{MP}\, g_{NQ}P_K\bar{P}^K\nonumber \\
&&
-{1\over 4}\, G_{MPC}\, \bar{G}_{NQ}^{\phantom{NQ}C}
-{1\over 4}\, \bar{G}_{MPC}\, G_{NQ}^{\phantom{NQ}C}\, ,\nonumber\\
(X_1)_{MNPQ} &\equiv& {1\over 2}\left( D_{M}G_{NPQ}+P_M\, \bar{G}_{NPQ}\right)
 + {1\over 4}
\left(
\bar{G}_{MNP}\, P_Q+\bar{G}_{PQM}\, P_{N}\right)\nonumber \\&& -{1\over 4}\, g_{MP}\, 
\bar{G}_{NQC} P^C\, ,\\
(X_2)_{MNPQ} &\equiv& - \, g_{MP}\, D_{N}P_Q-{1\over 4}\, G
_{MPC}\, G_{NQ}^{\phantom{NQ}C} \, ,\nonumber \\
(X_3)_{MNPQ} &\equiv& {1\over 4} \left( G_{MNP} P_Q + G_{PQM} P_N \right) -{1\over 4}\,
g_{MP} \, G_{NQC} P^C \, ,\nonumber\\
(X_4)_{MNPQ} &\equiv& -{3\over 4}\, 
g_{MP}P_N P_Q-{1\over 2}\, g_{MP}\, g_{NQ}P_CP^C\, . \nonumber 
\end{eqnarray}
In addition  we will denote
\begin{eqnarray}
\Big{(}t_8t_8+
\frac{1}{8}\varepsilon_{10}\varepsilon_{10}
\Big{)}X_i\,X_j\,X_k\,X_l &\equiv& 
\Big{(}t_8^{ABCDEFGH}t_8^{MNPQRSTU}+
\frac{1}{8}\varepsilon_{10}^{ABCDEFGHIJ}{\varepsilon_{10}^{MNPQRSTU}}_{IJ}
\Big{)} \nonumber  \\
&&
\times (X_i)_{ABMN}\, (X_j)_{CDPQ}\, (X_k)_{EFRS}\, (X_l)_{GHTU}\, ,
\end{eqnarray} 
for $i,j,k,l=0,...,4$ and similarly for interactions involving the complex 
conjugates $\bar{X}_i$.  
Then, the action takes the form
\begin{eqnarray}
{\cal{S}} &=& \frac{1}{2\kappa_{10}^2}\int d^{10}x \sqrt{-g} 
\left\{ R-2\, P_M\bar{P}^M - \frac{1}
{12}\, G_{MNP}\bar{G}^{MNP}\right. 
+\frac{{\alpha'}^3}{3\cdot 2^7}
\Big{(}t_8t_8+
\frac{1}{8}\varepsilon_{10}\varepsilon_{10}
\Big{)}\times\nonumber \\
\left[ {1\over 2}\right. f_0 \!\!\!\!\!\!\!\!\!\!\!\!\!&&\left( 12\,\bar{X}_1\,X_0^{2}\,X_1 \right.
+ 4\,\bar{X}_3\,X_1^{3} + 6\,\bar{X}_3^{2}\,X_3^{2} 
+ 12\,\bar{X}_3\,\bar{X}_1\,X_2^{2} + X_0^{4} \nonumber \\
 & & \mbox{} + 24\,\bar{X}_3\,\bar{X}_2\,X_2\,X_3 
+ 12\,\bar{X}_2^{2}\,X_1\,X_3 + 12\,\bar{X}_1^{2}
\,X_0\,X_2 + 12\,\bar{X}_3\,X_0^{2}\,
X_3 \nonumber \\
 & & \mbox{} + 24\,\bar{X}_2\,\bar{X}_1\,X_0\,X_3 
+ 24\,\bar{X}_1\,\bar{X}_2\,X_1\,X_2 + 24\,
\bar{X}_1\,\bar{X}_3\,X_1\,X_3 \nonumber \\
 & & \mbox{} + 24\,\bar{X}_3\,X_0\,X_1\,X_2 
+ 6\,\bar{X}_2^{2}\,X_2^{2} + 4\,\bar{X}_1^{3}\,
X_3 + 6\,\bar{X}_1^{2}\,X_1^{2}  \nonumber \\
 & & \mbox{} + 12\,\bar{X}_2\,X_0^{2}\,X_2 + 12\,
\bar{X}_2\,X_0\,X_1^{2} + 24\,\bar{X}_4\,\bar{X}_1\,X_1\,X_4 \nonumber \\
 & & \mbox{} + 24\,\bar{X}_4\,\bar{X}_1\,X_2
\,X_3 + 24\,\bar{X}_3\,X_4\,\bar{X}_1\,X_0 + 6\,\bar{X}_4^{2}\,X_4^{2} \nonumber \\
 & & \mbox{} + 24\,\bar{X}_4\,\bar{X}_3\,X_3\,X_4 + 12\,\bar{X}_2\,X_3^{2}
\,\bar{X}_4 + 12\,\bar{X}_3^{2}\,X_4\,X_2 + 12\,\bar{X}_1^{2}\,X_4
\,\bar{X}_2 \nonumber \\
 & & \mbox{} + 12\,\bar{X}_2^{2}\,X_0\,X_4 + 12\,X_2^{2}\,\bar{X}_4
\,X_0 + 12\,X_2\,\bar{X}_4\,X_1^{2} + 12\,X_0^{2}\,X_4\,\bar{X}_4 \nonumber \\
 & & \mbox{} + 24\,\bar{X}_3\,\bar{X}_2\,X_1\,X_4
+ \left. 24\,X_0\,X_1\,\bar{X}_4\,X_3 + 24\,
X_2\,\bar{X}_4\,X_4\,\bar{X}_2\right)\nonumber \\
+f_1 \!\!\!\!\!\!\!\!\!\!\!\!\!&&\left( 12\,\bar{X}_2\,X_1^{2}\,X_2\right. 
+ 12\,\bar{X}_2\,X_0\,X_2^{2} + 4\,\bar{X}_1
\,X_1^{3} + 6\,\bar{X}_1^{2}\,X_2^{2} \nonumber \\
 & & \mbox{} + 24\,\bar{X}_3\,X_0\,X_2\,X_3 
+ 12\,\bar{X}_3\,\bar{X}_1\,X_3^{2} + 24\,\bar{X}_1
\,\bar{X}_2\,X_2\,X_3 + 6\,\bar{X}_2^{2}\,
X_3^{2} \nonumber \\
 & & \mbox{} + 12\,\bar{X}_3\,X_1^{2}\,X_3 + 12\,
\bar{X}_3\,X_1\,X_2^{2} + 12\,\bar{X}_1^{2}\,
X_1\,X_3 + 12\,\bar{X}_1\,X_0^{2}\,X_3\nonumber \\
 & & \mbox{} + 24\,\bar{X}_2\,X_0\,X_1\,X_3 
+ 24\,\bar{X}_1\,X_0\,X_1\,X_2 + 4\,
X_2\,X_0^{3} + 6\,X_0^{2}\,X_1^{2} \nonumber \\
 & & \mbox{} + 12\,\bar{X}_4\,X_1^{2}\,X_4 + 6\,
X_4^{2}\,\bar{X}_3^{2} + 4\,X_2^{3}\,\bar{X}_4
 + 24\,X_3\,X_4\,\bar{X}_4\,\bar{X}_1 \nonumber \\
 & & \mbox{} + 24\,X_3\,X_4\,\bar{X}_3\,\bar{X}_2 
+ 24\,\bar{X}_2\,\bar{X}_1\,X_1\,X_4 + 12\,
\bar{X}_1^{2}\,X_4\,X_0 \nonumber \\
 & & \mbox{} + 12\,X_0\,X_3^{2}\,\bar{X}_4 + 12\,
X_4^{2}\,\bar{X}_4\,\bar{X}_2 + 12\,\bar{X}_2^{2}\,
X_2\,X_4 + 12\,X_0^{2}\,X_4\,\bar{X}_2 \nonumber \\
 & & \mbox{} + 24\,\bar{X}_3\,X_0\,X_1\,X_4 
+ 24\,\bar{X}_4\,X_1\,X_2\,X_3 + 24\,
\bar{X}_1\,X_4\,\bar{X}_3\,X_2 \nonumber \\
 & & \mbox{} + \left. 24\,\bar{X}_4\,X_4\,X_0\,X_2\right)\nonumber \\
+f_2 \!\!\!\!\!\!\!\!\!\!\!\!\!&&\left( 12\,\bar{X}_3\,X_2^{2}\,X_3 \right.
+ 12\,\bar{X}_2\,X_0\,X_3^{2} + 12\,\bar{X}_1
\,X_1\,X_2^{2} + 4\,\bar{X}_2\,X_2^{3} + X_1^{4} \nonumber \\
 & & \mbox{} + 24\,\bar{X}_1\,X_0\,X_2\,X_3 
+ 6\,\bar{X}_1^{2}\,X_3^{2} + 12\,\bar{X}_3\,X_1
\,X_3^{2} + 12\,\bar{X}_1\,X_1^{2}\,X_3\nonumber \\
 & & \mbox{} + 24\,\bar{X}_2\,X_1\,X_2\,X_3 
+ 6\,X_0^{2}\,X_2^{2} + 12\,X_0^{2}\,X_1
\,X_3 + 12\,X_0\,X_1^{2}\,X_2 \nonumber \\
 & & \mbox{} + 24\,\bar{X}_3\,X_1\,X_2\,X_4 
+ 12\,X_0\,X_4^{2}\,\bar{X}_4 + 12\,X_2^{2}
\,\bar{X}_4\,X_4 + 12\,\bar{X}_1^{2}\,X_2\,
X_4 \nonumber \\
 & & \mbox{} + 12\,X_2\,\bar{X}_4\,X_3^{2} + 12\,
\bar{X}_2\,X_1^{2}\,X_4 + 12\,X_4^{2}\,
\bar{X}_3\,\bar{X}_1 + 6\,X_4^{2}\,\bar{X}_2^{2} \nonumber \\
 & & \mbox{} + 4\,X_0^{3}\,X_4 + 24\,X_0\,
X_4\,\bar{X}_2\,X_2 + 24\,X_0\,X_4
\,\bar{X}_1\,X_1 + 24\,X_0\,X_4\,\bar{X}_3\,X_3 \nonumber \\
 & & \mbox{} + \left. 24\,X_1\,X_3\,\bar{X}_4\,X_4 + 24\,X_3\,
X_4\,\bar{X}_2\,\bar{X}_1\right)\nonumber \\
+f_3 \!\!\!\!\!\!\!\!\!\!\!\!\!&&\left( 12\,X_4^{2}\,\bar{X}_2\,X_0 + 12\,X_4^{2}\,
\bar{X}_3\,X_1 \right.+ 12\,\bar{X}_1\,X_2^{2}\,X_3 + 4\,\bar{X}_3\,X_3^{3} \nonumber \\
 & & \mbox{} + 12\,\bar{X}_2\,X_2\,X_3^{2} + 4\,
X_0\,X_2^{3} + 12\,\bar{X}_1\,X_1\,X_3^{2} 
+ 4\,X_3\,X_1^{3} + 6\,X_0^{2}\,
X_3^{2} \nonumber \\
 & & \mbox{} + 24\,X_1\,X_0\,X_2\,X_3
 + 6\,X_1^{2}\,X_2^{2} + 24\,\bar{X}_1\,X_4
\,X_0\,X_3 + 24\,\bar{X}_1\,X_1\,X_2\,X_4 \nonumber \\
 & & \mbox{} + 24\,\bar{X}_3\,X_3\,X_2\,X_4 
+ 12\,X_4^{2}\,X_2\,\bar{X}_4 + 12\,\bar{X}_2
\,X_2^{2}\,X_4 + 12\,\bar{X}_4\,X_4\,X_3^{2} \nonumber \\
 & & \mbox{} + \left. 12\,X_0^{2}\,X_2\,X_4 + 12\,X_1^{2}\,X_4\,X_0 + 6\,X_4^{2}\,
\bar{X}_1^{2} + 24\,X_3\,X_4\,\bar{X}_2\,X_1 \right)\nonumber \\
+f_4 \!\!\!\!\!\!\!\!\!\!\!\!\!&&\left( 4\,\bar{X}_1\,X_3^{3} + X_2^{4} \right.+ 
12\,X_0\,X_2\,X_3^{2} + 12\,X_1
\,X_2^{2}\,X_3 + 6\,X_1^{2}\,X_3^{2} \nonumber \\
 & & \mbox{} + 12\,X_3^{2}\,X_4\,\bar{X}_2 + 12\,
X_4^{2}\,\bar{X}_2\,X_2 + 6\,X_0^{2}\,
X_4^{2} + 12\,X_3\,X_4^{2}\,\bar{X}_3 \nonumber \\
 & & \mbox{} + 12\,X_0\,X_4\,X_2^{2} + 12\,
X_4^{2}\,\bar{X}_1\,X_1 + 12\,X_2\,X_4
\,X_1^{2} + 24\,X_3\,X_4\,\bar{X}_1\,
X_2 \nonumber \\
 & & \mbox{} + \left. 24\,X_3\,X_4\,X_0\,X_1
 + 4\,X_4^{3}\,\bar{X}_4\right)\nonumber \\
+f_5 \!\!\!\!\!\!\!\!\!\!\!\!\!&&\left( 4\,X_4^{3}\,\bar{X}_2 \right. + 12\,X_4^{2}
\,\bar{X}_1\,X_3 + 12\,X_4^{2}\,X_0
\,X_2 + 6\,X_4^{2}\,X_1^{2} \nonumber \\
 & & \mbox{} + 12\,X_0\,X_4\,X_3^{2} + 24\,
X_1\,X_3\,X_2\,X_4 + 4\,X_1\,
X_3^{3}\left. + 4\,X_2^{3}\,X_4 + 6\,X_2^{2}\,X_3^{2}\right)\nonumber \\
+f_6 \!\!\!\!\!\!\!\!\!\!\!\!\!&&\left( 4\,X_4^{3}\,X_0 + 12\,X_4^{2}
\,X_1\,X_3 + 6\,X_4^{2}\,X_2^{2} + 12
\,X_3^{2}\,X_4\,X_2 + X_3^{4}\right)\nonumber \\
+f_7\!\!\!\!\!\!\!\!\!\!\!\!\!&&\left( 4\,X_4^{3}\,X_2 + 6\,X_3^{2}\,
X_4^{2}\right)+f_8 \left( X_4^{4} \right) +\, \mbox{c.c.}.\left.\left.
\phantom{{1\over 2}}\!\!\!\!
 \right] \right\}\, .\label{inter}
\end{eqnarray}

\newpage
\begin{flushleft}
{\large\bf Appendix B}\end{flushleft}
\renewcommand{\theequation}{B.\arabic{equation}}
\renewcommand{\thesection}{B.}
\setcounter{equation}{0}

We present 
here some properties of the non-holomorphic $f_k$ functions.
They have been defined in eq.(\ref{fk}) as
\be
f_k(\tau,\btau)=
{\sum_{p,q}}'\frac{\tau_2^{3/2}}{(p\tau+q)^{3/2+k}(p\btau+q)^{3/2-k}}\, ,
\label{fkbis}
\ee
from which we derive the recursion formula for $k\in {\bf Z}$
\begin{eqnarray}
\left(k+2i\tau_2\frac{\p}{\p\tau}\right)f_k&=&\left(\frac{3}{2}+k\right)
f_{k+1} \, . \label{k1} 
\end{eqnarray}
By using the above relation, we find that their small 
$e^\phi=\tau_2^{-1}$ expansion is
\be
f_k(\tau,\btau)&=& 2\zeta(3)
\tau_2^{3/2}+{2\pi^2\over 3}c_k\tau_2^{-1/2}
+4\pi^{}\sum_{m\geq 1,\, n\geq 1}
{m^{1/2}\over n^{3/2}}\times \nonumber \\
&&\left[ \sum_{r=-k}^{\infty}
{C_{-k,r}\over (4\pi mn \tau_2)^{r}}
e^{2i\pi mn \tau}+\sum_{r=k}^{\infty}
{C_{k,r}\over (4\pi mn \tau_2)^{r}} e^{-2i\pi mn \btau}\right]\, , \label{exp}
\ee
where
\be
c_k& =&(-)^k\, {\pi\over 4}\, {1\over\Gamma(3/2+k)\Gamma(3/2-k)}\, , \nonumber
\\  r\geq l\in {\bf Z}\, ,~~C_{l,r}&=&{(-)^l\over (r-l)!}\, 
{\Gamma(3/2)\over \Gamma(l+3/2)}\, 
{\Gamma(r-1/2)\over \Gamma(-r-1/2)}\, .
\ee

An alternative form of the functions $f_k$ is 
\be
f_k(\tau,\btau)=\frac{4\pi^2}{4k^2-1}{\sum_{p,q}}'
\frac{\tau_2^{-1/2}}{(p\tau+q)^{-1/2+k}(p\btau+q)^{-1/2-k}}\, ,\label{alt}
\ee
which is a particular case of eq.(\ref{45}) for $s=3/2$.
In fact, the above sum does not converge but is defined by an analytic 
continuation \cite{M} similar to the well-known zeta-regularization. 

Using eq.(\ref{fkbis}), the  functions $f_k$ can be seen to satisfy the 
differential  equation
\be
(\tau-\btau)^2\p\bp f_k + k (\tau -\btau) \p f_k+ k (\tau -\btau) \bp f_k
+{3\over 4}f_k=0\, .
\label{diff}
\ee 
In fact, it is shown in \cite{M} that any modular function 
of holomorphic and anti-holomorphic weights $({3\over2}+k,{3\over 2}-k)$, 
 which satisfies  the above  equation,
must be  equal to $f_k$.

\end{document}